# Inflation targeting strategy and its credibility

Carlos Esteban POSADA[1]

### Abstract

*The money supply is endogenous if the monetary policy strategy is the so-called "Inflation and Interest Rate Targeting" (IRT). With that and perfect credibility, the theory of the price level and inflation only needs the Fisher equation, but it interprets causality in a new sense: if the monetary authority raises the policy rate, it will raise the inflation target, and vice versa, given the natural interest rate. If credibility is not perfect or if expectations are not completely rational, the theory needs something more. Here I present a model corresponding to this theory that includes both the steady state case and the recovery dynamics after a supply shock, with and without policy reactions to such a shock. But, under the finite horizon assumption for IRT, at some future point in time the money supply must become exogenous. This creates the incentive for agents to examine, as of today, statistics on monetary aggregates and form their forecasts of money supply growth and inflation rates. Additionally, inflation models of the small open economy allow us to deduce that the IRT in this case is much more powerful than otherwise, and for the same degree of credibility. But things are not necessarily easier for the monetary authority: it must monitor not only internal indicators, but also external inflation and its determinants, and it must, in certain circumstances, make more intense adjustments to the interest rate.*

---

[1] Professor, School of Finance, Economics and Government, Universidad EAFIT (Colombia). Address: cposad25@eafit.edu.co.

**I. Introduction**

This paper deals with the meaning and some implications of the "Inflation and Interest Rate Targeting Strategy" (*IRT*). The first thing, then, is to clarify that *IRT* is not a remedy to achieve an inflation target. It is simply a strategy to "apply the cure", and the validity and relevance of the strategy is conditioned on several factors, including the wisdom and willingness of the monetary authority to select and apply the remedy, and the credibility it enjoys among private agents about its ability to hit the target.

But that brings us directly to an important point in the current discussions: that of "dominance". Is the monetary authority dominated by the fiscal authority? That is, are the evolution of fiscal variables (tax revenues/GDP, public expenditures and subsidies/GDP, public debt/GDP, etc.) and, ultimately, the decisions taken in this respect by the fiscal authority the basic determinants of an economy's monetary flows? Is it, then, a certain economy characterized by "fiscal dominance"? If the answer is yes, what is relevant to interpret inflationary trends and cycles is the family of fiscal theory models of the price level and inflation (Bianchi et al. 2022). Indeed, if the role of the fiscal authority is dominant, it can be assumed that changes in public expenditures and transfers to the private sector or changes in tax policy sooner or later lead to changes in the money supply, policy interest rates, and inflation targets and expectations[2].

Regarding the theoretical origin of *IRT*, the work of William Poole (1970) should be highlighted, who considered the instability of money demand (in his opinion significantly larger than that of many other relevants macroeconomic variables) brought about a problem for the implementation of the monetarist strategy: fixing the rate of growth of the money supply generated high volatility in interest rates (and, for economies more open than the US, such as those of Western Europe, excessive exchange rate fluctuations arising from interest rate gyrations).

Cochrane (2022) presents a model for interpreting the scope and assumptions of the *IRT*; his model is neo-Keynesian with rational expectations and an exogenous fiscal component (the change in the expected present value of the sum of future fiscal surpluses, one of the determinants of the change in inflation expectations). Cochrane clarifies that, under different versions, his model may give different results on the evolution of inflation in the long run following a temporary reduction in inflation. Inflation may fall by a certain magnitude now but subsequently rise by a large amount, as Sargent and Wallace (1981) had already shown by illustrating a case of inconsistency between monetary and fiscal policies.

In this paper I consider only the case of dominance of the monetary authority: dominance in the sense that it takes decisions autonomously to the point of inducing the fiscal authority to modify tax, expenditure, and subsidy policies to minimize the eventual social costs of an inconsistency between monetary and fiscal policies. Thus, in what follows it is assumed that there is not an exogenous fiscal variable that could be important in the determination of the inflation target and inflation expectations.

---

[2] The fiscal theory of inflation assumes fiscal dominance. But it has an implicit assumption something weird: that the monetary authority is <u>distinct from the fiscal authority but subordinate to it.</u> This may be true but, in practice, it is an unstable situation: it may end up in a situation of very high inflation or hyperinflation or, on the contrary, in an excessive deflation.

The objective of this paper is, then, to set out as simply as possible the theory underlying the *IRT* and the credibility it may enjoy under the premise of monetary authority dominance. To achieve its objective the paper has, in addition to this introduction, the following sections: II. Infinite horizon: the steady state, and the dynamics generated from an inflation target violation; III. Finite horizon; IV. The case of the small open economy; V. Conclusions.

This paper does not discuss optimal rules for the society. It is considered that the central banker is subject to one rule: that observed inflation is equal to the target; that is, it is assumed that the only objective of the central banker is to succeed in reaching the target.

## II. Infinite horizon: the steady state and the dynamics generated from an inflation target violation

The starting point is the application of Fisher's hypothesis to the policy interest rate in a steady state path:

$$\bar{R} = (1 + r^{ss})(1 + E_t \pi_{t+1}) - 1$$

$\bar{R}$, $r^{ss}$, $E_t \pi_{t+1}$ being the nominal policy interest rate (set by the monetary authority), the steady-state real interest rate (or natural rate; we assume it is an exogenous constant), and the expected inflation rate in period t for the following period. Here, and throughout this section, we will assume that the inflation expectation is equal to the inflation target set by the monetary authority and equal to the inflation rate observed in each of the t+x periods, for *x = 1, 2, ...,* over a horizon extending to infinity.

From the above equation (and assuming perfect credibility) it follows that:

$$(II.1) \ \ E_t \pi_{t+1} = \frac{1 + \bar{R}}{1 + r^{ss}} - 1$$

But the definition of the expected rate of inflation is as follows:

$$E_t \pi_{t+1} = \frac{E_t P_{t+1}}{P_t} - 1$$

Where *P* is the general price level. Therefore:

$$P_t = \frac{E_t P_{t+1}}{1 + E_t \pi_{t+1}}; \ \ E_t P_{t+1} = \frac{E_{t+1} P_{t+2}}{1 + E_{t+1} \pi_{t+2}}; \dots;$$

$$(II.2) \ \ E_{t+x} P_{t+x+1} = \frac{E_{t+x+1} P_{t+x+2}}{1 + E_{t+x+1} \pi_{t+x+2}} \dots$$

and so on, for $x \to \infty$.

## 1. Stability: Missing the target and the return to the steady state

Is such a path credible, and is it stable? These questions amount to trying to know what might happen in the event of an (unforeseen) shock to the price level generating a discrepancy between the observed rate of inflation and the target. If such a discrepancy is transitory, that is, if it is exhausted over time, and the (high) speed of the transition can be attributed, at least in part, to the confidence of private agents in the monetary policy, then the monetary authority will preserve credibility in its strategy despite the fact that the forecast will only be correct for the long term, taking into account that, in reality, a shock may be followed by others (some positive and some negative shocks, all unforeseen).

The statements in the previous paragraph assume several things, but one of them is that individual prices and the general price level of the economy are not (instantaneously) flexible, that is, they do not adjust instantaneously to shocks that could give rise to markets imbalances. Therefore, the relevant model to defend such assertions is the sticky prices New Keynesian one (whose adjustment is slow but, in the long run, complete). Another assumption of what is stated in the previous paragraph is that the monetary authority follows a rule of conduct: it tries to ensure that observed inflation is equal to the target (in a not too long term), and that, if a discrepancy arises in this regard, it will do everything in its power to eliminate said discrepancy. Adhering to such a rule is fundamental for agents to give credibility to the monetary authority's pronouncements, forecasts, and actions.

The New Keynesian model accompanied by that rule (which, unlike the Taylor rule, omits another of the two possible concerns of the monetary authority, that is, trying to minimize a difference between observed and optimal output) is made up of three equations under its structural form, namely[3]:

$$\pi_t = \vartheta + \beta E_t \pi_{t+1} + \gamma x_t + \epsilon_{\pi t}; \ \vartheta, \gamma > 0, \beta \equiv \frac{1}{1+\theta}, 0 < \theta < 1$$

$$x_t = E_t x_{t+1} - \alpha(R_t - E_t \pi_{t+1} - \theta) + \epsilon_{xt}; \ \alpha > 0$$

$$R_t = \theta + \pi^* + \mu(\pi_t - \pi^*); \ \mu > 0$$

Where: $x, \theta, \pi^*, \epsilon_{i(i=\pi,x)}$ are the output gap (or difference between observed and optimal or steady-state GDP), the subjective discount rate, the inflation target, and the unanticipated inflation and output shocks, respectively, and $\vartheta, \gamma, \alpha, \mu$ are parameters.

The solution of the model for the steady state situation is:

$$\pi_t = \pi^*; \ x_t = 0; \ R_t = \bar{R} = \theta + \pi^*$$

In this model, the natural rate of interest is equal to the subjective discount rate as the assumption that output per worker grows in the long run is omitted.

---

[3] Wickens (2011; pp. 413 ff.) presents an almost identical version of this model. The main difference is that this author includes the Taylor rule and considers the possibility of (unforeseen) interest rate shocks.

The dynamic properties of the model are tested by means of simulation exercises with computer programs since there are no unique and determinate analytical solutions (and robust to wide ranges of numerical values of the parameters) to describe the transition generated by a shock giving rise to output and inflation gaps and subsequently leads to their gradual disappearance until returning to the steady state.

The following is a simple exercise with an analytical solution that replicates what the New Keynesian model predicts with the inflation rule we have been contemplating and illustrates what credibility means.

Suppose that a sufficiently large portion of individual prices is of the sticky type; in such a case, the overall price level will also be sticky. Thus, if a shock consists of something inducing a transitory gap between observed real GDP ($Y$) and optimal or steady-state GDP ($Y^*$), either a demand shock or a supply shock, then we can deduce there will be a variation in the price level over time, such that only slowly will the gap be eliminated, with such a price variation being additional to the inflation rate we could observe without such a shock. In more precise terms, let us express such variation of the price level (abstracting from inflation for the time being; this is why the lower-case letter is used for this price level) so:

$$(II.3) \quad \frac{dp_t}{dt} = f(Y - Y^*); \;\; f(0) = 0; f' > 0$$

It should be clarified that equation II.3 refers only to the change in the price level that begins to be observed from the moment immediately after the shock, i.e., what corresponds to the "sticky" change, since at the time of the shock it can occur an instantaneous jump in the price level. In other words, the "sticky" price´s characteristic is associated with the normal functioning of certain markets and does not exclude the possibility that the same product´s price exhibits, in certain circumstances, an instantaneous jump and, in other circumstances, typical of what could be called normality in its market, its variations will be gradual and only slowly contribute to reestablish the market equilibrium.

The simplest case is the following:

$$Y = b_1 - a_1 p; \;\; a_1 > 0;$$

$$Y^* = b_2; \;\; b_1 > b_2$$

Equilibrium implies that:

$$Y = Y^*$$

$$\therefore$$

$$(II.4) \;\; p^* = \frac{c_1}{a_1}; \; c_1 \equiv b_1 - b_2$$

In such a case, the hypothesis (II.3) can be applied in a simplified form, thus:

$$(II.5) \quad \frac{dp_t}{dt} = q(b_1 - a_1 p_t - b_2); \quad q > 0$$

$$\therefore \ \frac{dp_t}{dt} + q a_1 p_t = q c_1$$

The above is a first order differential equation; its solution for the "homogeneous equation" case (i.e., omitting the constant) is:

$$p_t = e^{-q a_1 t}$$

Such an equation expresses the law of evolution of the price level for an initial level (*t = 0*) equal to 1, and for a transitory shock equal to an arbitrary value.

Therefore, the general solution (already including the constant) and with a transitory shock $s$ is:

$$(II.6) \ \ p_t = p_t^* + s e^{-q a_1 t}$$

In words, the general price level in the period of the shock takes the value $p_t^* + s e^{-q a_1 t}$ , but, from the following moment, the difference between the observed level and the equilibrium level, $p_t^*$, tends to reduce until its elimination, unless a new shock occurs[4].

In posing equation II.3 it was mentioned that inflation is abstracted upon that. Now we must face this issue. To proceed without further complications, we will assume that $p_t$ is equal to the general level of monetary prices, including the effects of the shock (in the period in which it occurs and in the following periods), deflated by the expected rate of inflation, which is equal, by hypothesis, to the inflation target (the case of total credibility), this being an exogenous variable:

$$(II.7) \ \ p_t = \frac{P_t}{1 + \pi^*}$$

Thus, $p_t$ is the "real" price of output, and so it makes sense to take it into account in the exercise leading to condition II.4 (agents, supposedly, make decisions on prices and quantities of output considering the expected rate of inflation, which is equal to the target). But the general price level receives the impact of the shock, so the observed inflation rate must differ transiently from the inflation target:

$$\pi_t = \frac{P_{t+1} - P_t}{P_t}$$

---

[4] Nothing substantial would change if it is assumed that $Y^* = b_2 + a_2 p$; $a_2 \gtreqless 0$, provided that $a_1 + a_2 > 0$. With such a condition (and with $b_1 - b_2 > 0$) the requirement of system stability is satisfied and, therefore, hypothesis II.3 is respected.

Therefore, only in the steady state path (i.e., for $t = ee$) it must be fulfilled that: $p_{ee} = P_{ee}/1 + \pi_{ee}$

The above may have a limitation: the rate of change of the "real" price would appear to be different in economies with very low or no inflation *vis-á-vis* economies with high inflation (Alvarez and Lippi 2020); so, a better model should incorporate different sensitivities of those rates to different ranges of inflation. This means that, in our case, the parameters $a_1, q$ would have values conditional on a certain relatively narrow inflation rates´ range.

**Transition dynamics and the other policy task**

The process of adjusting inflation to its steady state level in the face of (for example) a negative supply shock (which causes a positive price shock) may be too slow, and this could undermine the credibility of the inflation targeting strategy. A second task of monetary policy is to reduce the time required for such a process (the first task is to set the target and the policy interest rate consistent with hitting the target).

This second task is accomplished by the monetary authority setting a transitory "premium" (positive or negative) on the steady state policy interest rate, thus:

$$(II.8) \ \rho_t = 1 + R_t - \bar{R};$$

$$\bar{R} = \theta + \pi^*$$

So, a positive premium ($R_t - \bar{R} > 0$) would have a negative effect on the price level and the inflation rate (and additional effects on the output gap). Therefore, the example should be extended to take this into account considering the factor $q$ may change over time as it depends on that premium, thus:

$$(II.9) \ q_t = j/\rho_t; \ j > 0$$

$$\Rightarrow$$

$$(II.9.a) \ q_t = j/(1 + R_t - \bar{R})$$

This means the sensitivity of the change in the price level to the imbalance between observed and optimal GDP depends on the premium.

**A numerical example**

A numerical exercise applying the above example is now presented. Table 1 contains the set of exogenous variables and parameters of the baseline scenario. This scenario is that of a monetary policy not varying in the face of a transitory and unforeseen positive shock to the price level; that is, the premium is always 1: the policy rate is always equal to the rate compatible with meeting the inflation target.

| Table 1. Baseline Scenario | | | | | | | |
|---|---|---|---|---|---|---|---|
| Inflation Target | $r^{ss}$ | $p^*$ in the begining | $b_1$ | $b_2$ | $a_1$ | $j$ | $\bar{R}$ |
| 0,03 | 0,015 | 1 | 1 | 0,85 | 0,15 | 0,6 | 0,0455 |

Figure 1 shows one result of the exercise in the baseline scenario, namely the time paths of the price level. One of the trajectories is that of the steady state, that is, without unforeseen shocks; the other trajectory corresponds to the price level when there is a one-time unforeseen shock to the price level equivalent to almost 10% of the magnitude of its previous level in a given period (period 20 was arbitrarily chosen); a transitory divergence is created, and then the effective price level starts to converge to the steady state level.

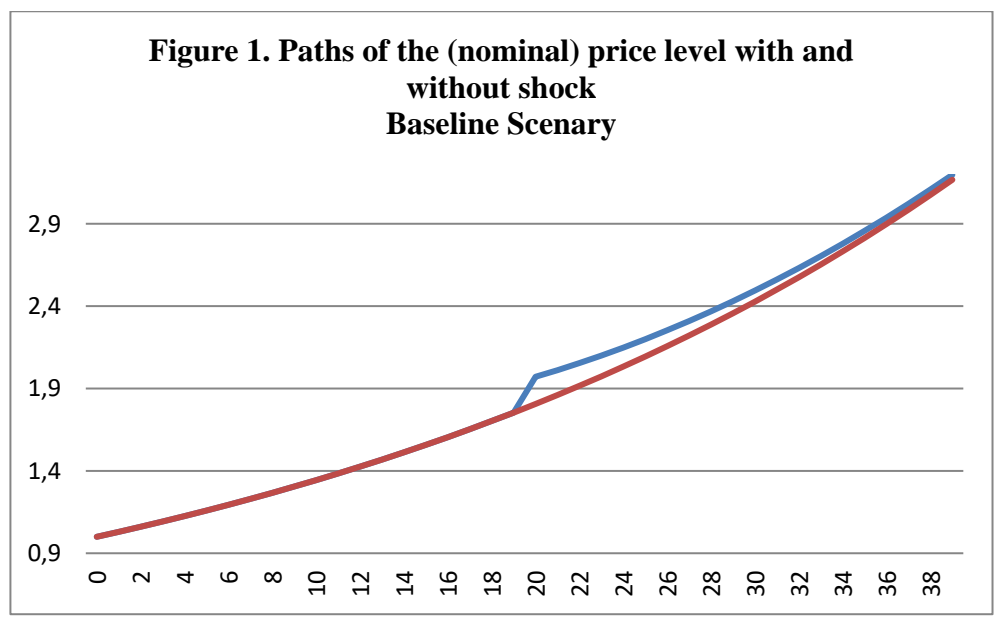

Figure 2 shows the trajectories of the inflation and output gap (the difference between observed and optimal output). The movements during the two periods immediately following the shock are of the "anti-Phillips curve" type, but the following movements are of the "Phillips curve" type (it should be remembered that the output gap depends on the "real" price or price level deflated with the inflation target; if it were deflated with the observed rate of inflation, the magnitudes of the output gap and its variations would be different).

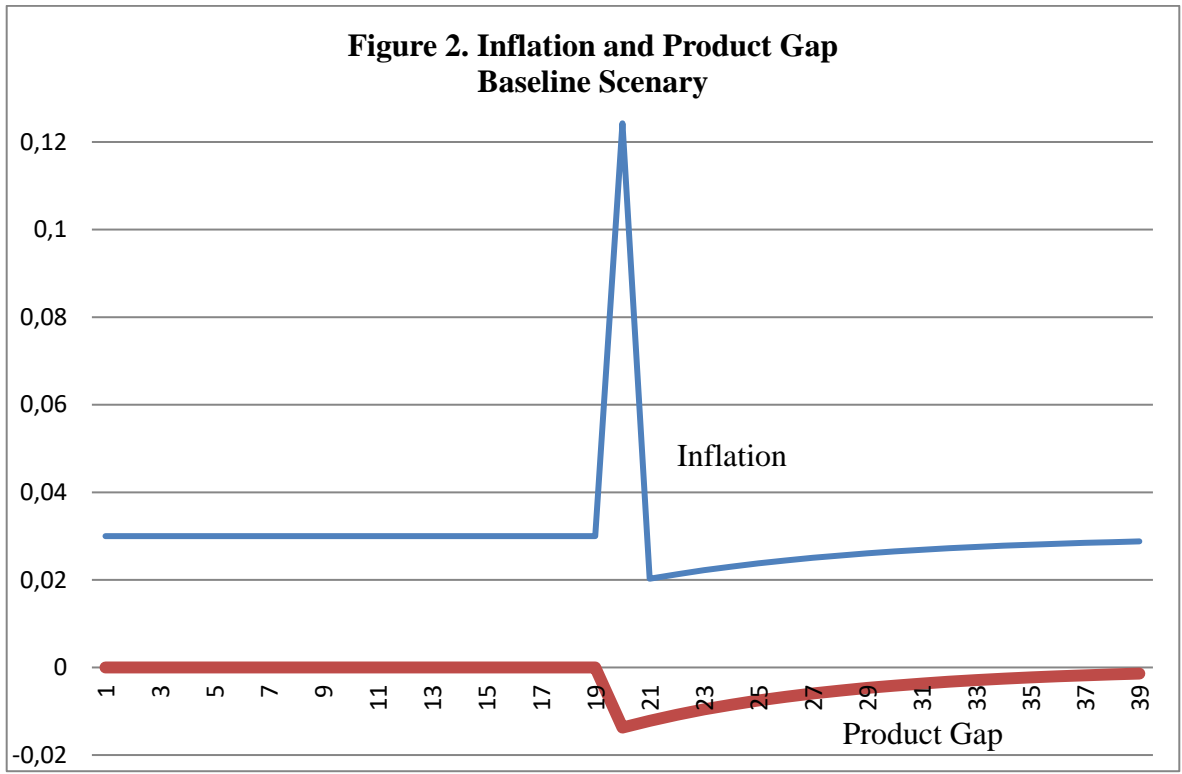

Figure 2. Inflation and Product Gap
Baseline Scenary

An alternative scenario is that of a restrictive policy implemented from the post-shock period since it is unforeseen. The policy rate is set at 9% in the period immediately following the shock, and in the 8 subsequent periods it is reduced by half a percentage point until it returns to the pre-shock level.

Figure 3 shows the inflation trajectories in both scenarios (baseline and with policy) between the time of the shock and the return to the steady state path. As can be seen in Figure 3, the effect of the policy (by setting the rate at a sufficiently high level, 9%, in the first period after the shock) is to make inflation almost zero in the period after the shock, but then inflation begins to rise, exceeding the baseline scenario´s inflation in some periods until it ends up being equal to of the baseline scenario and then converging to the target.

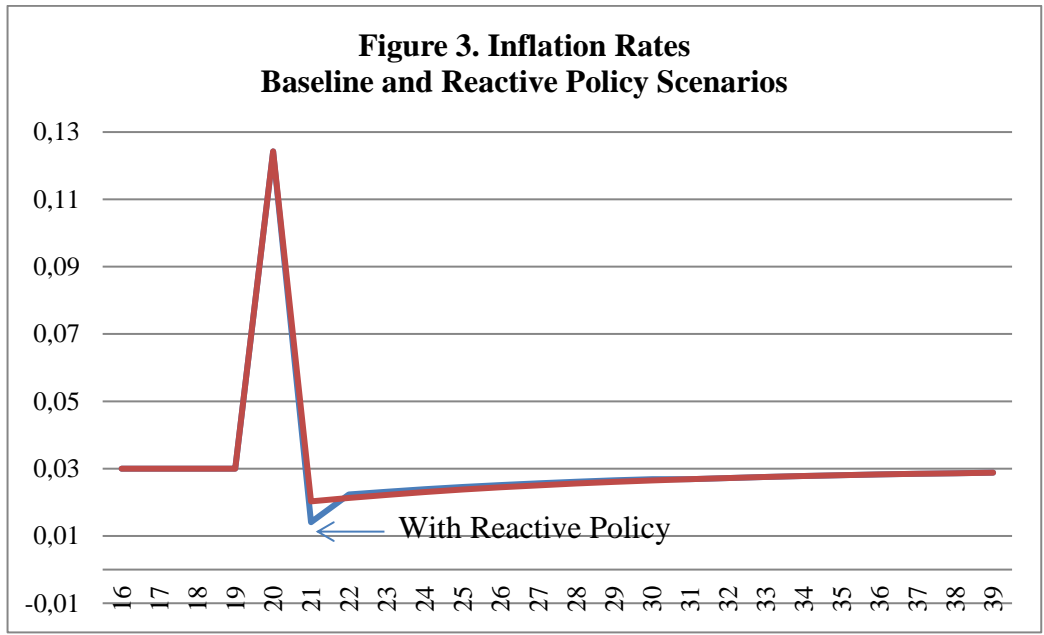

As an effect of reactive policy, the output gap in the reactive policy scenario eventually becomes slightly smaller than in the reference scenario (Figure 4), but the difference is negligible.

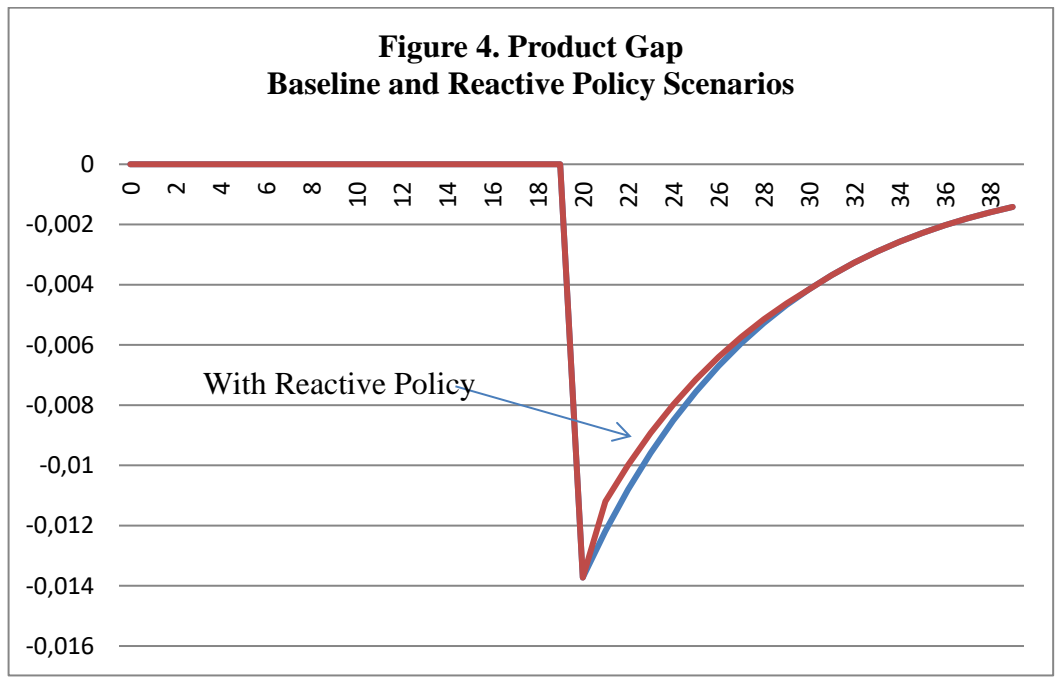

This can be summed up in one sentence: when there is (total) credibility in the monetary authority (and if the shock does not weaken it), it is of little importance to implement a reactive policy to counteract the transitory effects of price level shocks: the dynamics of the transition it is almost the same with or without countercyclical policy. This leads to the conclusion that the importance of a reactive monetary policy derives from the low (or zero) credibility of a monetary authority, and the perception that executing it contributes to recovering credibility.

## 2. Money: its demand and the endogeneity supply´s

Theory tells us the general equilibrium money price level is positively associated with the nominal money supply, $M^s$, and negatively associated with the equilibrium demand for real money balances, $L$, thus:

$$(II.10) \quad P_{t+x-1}^* = \frac{M_{t+x-1}^s}{L_{t+x-1}^*}; \quad L_{t+x-1}^* = L(Y_{t+x-1}^*, R_{t+x}); \; L_Y > 0; \; L_R < 0;$$

The contrast between equation II.10 and the steady state description (equations II.1 and II.2) plus the case described by equations II.3 to II.9 allows us to deduce that the *IRT* turns the money supply into an endogenous variable: it does not explain the price level or inflation; on the contrary, it is explained by the price level and the demand for real money balances, given the interest rate (which we assume determined by the monetary authority). In addition, this contrast shows the origin of the credibility of the monetary authority, namely: its ability to make the observed inflation rate converge to the target in a not very long period of time and, where appropriate, temporarily set an interest rate above or below the rate compatible with meeting the target.

## III. A finite horizon

But now suppose the horizon of the inflation targeting strategy is finite, that is, that it ends in period $t+T$ because private agents anticipate it for some reason. In such a case, the rational forecasts made in period t for the price level and the inflation rate at the end of the horizon can only be the following:

$$(III.1) \quad E_{t+T-1} P_{t+T} = \frac{E_{t+T-1} \bar{M}_{t+T}^s}{E_{t+T-1} L_{t+T}};$$

Therefore:

$$(III.2) \quad E_{t+x-1} \pi_{t+x} = \frac{E_{t+x-1} P_{t+x}}{E_{t+x-2} P_{t+x-1}} - 1, \quad x = 0, 1, \dots T$$

Indeed, when the horizon is finite (and ends in period $t+T$) it is impossible to apply equation II.1; the only possible thing is to consider that the variables on the right side of III.1 refer to the same period (the last one) of the variable on the left side (or to an earlier one). That is, at the end of the horizon foreseen for the *IRT* the price level can only be understood as determined by the contemporaneous values of real money demand, $L$, and of some exogenous variable. Therefore, what we must consider as an explanatory variable, in addition to $L$, is an exogenous nominal money supply, $M^s$ [5]

Applying III.1 to previous periods ("backward") it turns out:

$$(III.3) \;\; E_{t+x-1}P_{t+x} = \frac{E_{t+x-1}\bar{M}^s_{t+x}}{E_{t+x-1}L_{t+x}}; \;\; x = 0, 1, \dots T-1$$

$$\dots$$

$$P_t = \frac{\bar{M}^s_t}{L_t}$$

But this is compatible with the *IRT* such that:

$$E_{t+x-1}P_{t+x} = \frac{E_{t+x}P_{t+x+1}}{1 + E_{t+x-1}\pi_{t+x}}; \;\; x = 1, 2, \dots, T-1$$

The big difference with respect to the case of the infinite horizon is that, now, private agents will be careful to compare the time trajectories of the price level and inflation described in II.1 and II.2 with that described in III.3, that is that is, with the trajectories of the money supply and of the variables that determine the real demand for money. In other words, in the case of a finite horizon for the *IRT* strategy, even though it endogenizes the present money supply and that of some future periods, private agents will have incentives to make forecasts about the future values of the money supply and about what they imply in terms of long-term money supply growth rates because they must consider that this will be important for future inflation rates, and these, in turn, are likely to be important for the current one.

## IV. The case of the small open economy

This case is not substantially different from that of the closed economy, so we need not repeat what has been said above. However, there are two issues that deserve mention: a) this case makes the discussion of infinite versus finite horizons unnecessary, given the close and contemporaneous (or almost contemporaneous) connection between local and external inflation (the latter is considered an exogenous variable), as will be made explicit below; b) in certain circumstances, the monetary authority of the small open economy (*SOE*) will find it easier to hit the inflation target, and, in other circumstances, its task will be more difficult, if we compare its case with a much less open, such as that of the United States. This can be clarified with the following simplifications for the *SOE* case.

Clarifications and simplifications are presented using a model (as simplified as possible) to explain the determination of the *SOE* inflation as a process associated with the monetary policy executed in said economy, if it has its own currency (which we can call "peso"), and a flexible exchange rate.

---

[5] It is worth clarifying that if the horizon is finite and, for example, covers the next 10 years, the demand for real money balances foreseen today for the tenth year will depend on the forecast of the interest rate between year 10 and year 11, and this, by hypothesis, will no longer be the policy rate but the one set by the market.

$$(IV.1)\ \ P = \lambda P^* \text{T};$$

$$(IV.2)\ \ \text{T} = \gamma_0 e^{\gamma_1 (R - R^*)^{un}};\ \gamma_0 > 0;\ \gamma_1 < 0;$$

In equation IV.1 $P, \lambda, P^*, T$ are the nominal price level of the *SOE*, the inverse of the real exchange rate, the external (nominal) price level and the nominal exchange rate (pesos/unit of external currency), respectively. It is assumed that the real exchange rate and the external price level are exogenous variables, and the real exchange rate is constant.

Equation IV.2 is a huge simplification. It only refers to what has to do with our subject, that of the strategy of setting the policy interest rate required to hit the inflation target. What this equation postulates is that the exchange rate depends negatively on the unanticipated component, *un*, of the spread between domestic and foreign policy interest rates, $(R - R^*)^{un}$. This component, once observed, gives rise to arbitrage (which quickly runs out) that has a negative (permanent and definitive) effect on the nominal exchange rate. (i.e., a positive effect on the external value of the peso). In the absence of (unanticipated) policy innovations ("surprises"), i.e., in the case where exchange rate movements are explained by expectations that are always confirmed, policy decisions are not relevant to explain the *SOE* inflation. In other words, everything past and everything expected today is already incorporated in the exchange rate component called $\gamma_0$ [6]

Therefore:

$$(IV.2)\ en\ (IV.1) \Rightarrow$$

$$(IV.1.a)\ \ P = \lambda P^* \gamma e^{\gamma_1 (R - R^*)^{un}}$$

$$\Rightarrow$$

$$lnP = ln\lambda + lnP^* + ln\gamma_0 + \gamma_1 (R - R^*)^{un}$$

$$\Rightarrow$$

$$\frac{dlnP}{dt} = \frac{dlnP^*}{dt} + \frac{dln\gamma_0}{dt} + \gamma_1 \frac{d(R - R^*)^{un}}{dt}$$

That is:

:

$$(IV.3)\ \ \pi = \pi^* + \widehat{\gamma_0} + \gamma_1 \frac{d(R - R^*)^{un}}{dt}$$

So:

---

[6] The paper by Inoue and Rossi (2018) seems of great relevance related to the basic assumptions in this paragraph and to support equation IV.2, although it emphasizes the risk factor, and reports empirical evidence for the United States and other developed economies.

$$\frac{\partial \pi}{\partial (R - R^*)^{un}} < 0$$

According to IV.3, external circumstances (that is, those factors that affect the evolution of external inflation and the depreciation of the peso) can facilitate or hinder the task of the local monetary authority: they will require less or no monetary policy "innovation" or, conversely, more from it to keep inflation at a level equal to the target.

It should be clarified that, in the absence of innovations in the gap between domestic and foreign interest rates, and assuming exhausted arbitrage in the debt and exchange markets, the interest rate differential does not depend on the exchange rate; it is simply positively associated with the expected rate of depreciation of the local currency (the case of perfect capital mobility). This can be expressed as follows[7]:

$$\widehat{\gamma_0} \approx ln(E_t \mathrm{T}_{t+1}) - ln(\mathrm{T}_t) = ln(1 + E_t R_{t+1}) - ln(1 + E_t R^*_{t+1}) + \varepsilon_t$$

where $\varepsilon_t$ is a random component.

Another way of saying the above is this: a monetary policy always equal to what agents expect, if we assume that expectations are rational (and well-informed), is always incorporated in the expected rate of the peso depreciation, $\widehat{\gamma_0}$; observed inflation incorporates, sooner or later, this expectation.

## V. Conclusions

Under the assumptions of credibility of the monetary authority and infinite horizon for the *IRT*, the theory that supports and "rationalizes" this strategy has been called "New Fisherian", recalling Irving Fisher's hypothesis (García-Schmidt and Woodford 2019; Cochrane 2022) and summarized in equation II.1. With such hypotheses this equation expresses the following: the (actual and expected) inflation rate is positively associated with the (nominal) policy interest rate and negatively associated with the natural interest rate (the real rate prevailing in a steady state path). Furthermore, following this theory, the occurrence of unforeseen supply or demand shocks able of temporarily altering inflation could make it unnecessary to vary the policy interest rate: the credibility enjoyed by the monetary authority allows the inflation rate to return quickly to the inflation target without changes in the monetary policy. The results of numerical exercises presented in four figures in Section II are consistent with this hypothesis.

But going further to consider cases of imperfect credibility, namely, the risk of loss of credibility in the face of inflationary shocks or expectational errors, the IRT should consider (and actually does) the possibility of setting short-term policy rates that differ from the steady state policy rate so when the inflation rate rises (and it can be reasonably judged that its return to the target will not be fast enough) the policy rate will be increased in order to reduce observed inflation until it becomes evident that it will soon return to its target.

---

[7] See, e.g., Wickens (2011, ch. 13).

And if private agents judge that the *IRT* horizon is not infinite, it should not be surprising that they and the economists, and central banks pay attention to the money supply statistical series.

In the case of the small open economy, the impacts of a higher external inflation and of all those exogenous factors that increase the expected depreciation of the local currency will increase the inflation rate (through various channels). The *SOE´s IRT* is efficient in reducing the inflation rate to the extent that the monetary authority sets or increases unexpectedly the gap between the policy interest rate and the external interest rate. By doing so, and by declaring her willingness to repeat such action if she sees fit, she will succeed in reducing the nominal exchange rate and the expected peso´s depreciation; therefore, sooner or later the observed inflation rate will hit the target. And if the weight of non-rational expectations is relatively high, the monetary authority's decision to raise the interest rate will be efficient in reducing inflation even if there are several agents who have already anticipated this and acted accordingly.